\documentclass[a4paper,10pt]{article}
\usepackage{graphicx}

\usepackage{amssymb,amsmath}
\usepackage{textcomp}
\def\pacs#1{\vspace{10pt} \hspace{0.33cm} \rm PACS numbers: #1 \par \vspace{10pt}}

\title{Self-consistency  in non-extensive thermodynamics of highly excited hadronic states}
\author{A. Deppman}
\date{Instituto de F\'isica -  Universidade de S\~ao Paulo \\ email: deppman@if.usp.br \\ Rua do Matão Travessa R Nr.187 CEP 05508-090 Cidade Universitária, São Paulo - Brasil}

\begin{document}

\maketitle

\begin{abstract} The self-consistency of a thermodynamical theory for hadronic systems based on the non-extensive statistics is investigated. We show that it is possible to obtain a self-consistent theory according to the asymptotic bootstrap principle if the mass spectrum and the energy density increase q-exponentially. A  direct consequence is the existence of a limiting effective temperature for the hadronic system. We show that this result is in agreement with experiments.
\end{abstract}


The asymptotic bootstrap principle, which leads to a self-consistent thermodynamical theory of strong interactions at high energies, was a sucessful framework to understand high energy collisions. The theory  predicted a limiting temperature for the fireball produced during the reaction, and provided formulae for transverse momentum distributions  and for the mass spectrum of hadrons. The theory, fully developed by Hagedorn~\cite{Hagedorn}, was able to describe experimental data for center-of-mass (CMS) energies up to $\sqrt{s} \approx 20 GeV$. At higher energies, however, no agreement was found between calculations and experiments.

The solution was proposed independently by Bediaga~\cite{Bediaga2000156} and Beck~\cite{Beck} by including the so-called non-extensive statistics due to Tsallis~\cite{Tsallis1998} into the thermodynamical description of fireballs. This was done by the substitution of the Boltzmann factor which appears in the Hagedorn's theory by a generalization which takes the q-exponential function as a main ingredient, i.e.,
\begin{equation}
\exp\{-\beta E\} \rightarrow [1+(q-1)\beta E]^{-\frac{q}{(q-1)}}
\label{q-Boltzmann-factor}
\end{equation}
where $q$, called entropic factor, is an unknown quantity that is characteristic of the system.

With this simple substitution it was possible to recover the good agreement between calculation and experiment, even at energies as high as those achieved at LHC~ \cite{Chinellato2010, Alice20111655}, with $q>1$. The success of the generalized formalism indicates that there are correlations among partons in the quark-gluon plasma (QGP)~\cite{Kodama2005439}. These correlations generate temperature fluctuations in the non-equilibrated system~\cite{Wilk2007, Wilk2009}, which slowly moves through quasi-static states to an equilibrium point~\cite{Kodama2009289}. In fact, $\beta$ is the inverse of an effective temperature in the generalized formalism, although here we may refer to it simply as temperature.

In the present work we assume that the non-extensive thermodynamics is the correct framework for describing highly excited hadronic systems. We show that  in this framework it is possible to obtain a self-consistent theory of fireballs based on a generalization of the asymptotic bootstrap principle. We obtain new consistent mass spectrum and density of states, and show that there is a limiting temperature for the excited system.

Hagedorn's theory is based on the equivalence of two diferent forms for the partition function of excited hadronic systems,
\begin{equation}
 Z(V_o,T)=\int_0^{\infty}\sigma(E)\exp\{-\beta E\} dE
 \end{equation}
 and
 \begin{equation}
  \ln[1+Z(V_o,T)]=\frac{V_o}{2\pi^2}\sum_{n=1}^{\infty}\frac{1}{n}\int_0^{\infty}dm \int_0^{\infty}dp \, p^2 \rho(n;m)\exp\{-n\beta \sqrt{p^2+m^2}\} \,,
\end{equation}
where
\begin{equation}
\rho(n;m)= \rho _B(m) - (-1)^n \rho_F(m)
\end{equation}
with $\rho _B(m)$ and $\rho_F(m)$ being, respectively, the mass spectrum for bosons and for fermions, and $\sigma(E)$ being the density of states for the system at energy $E$.

The consistency between the two forms of the partition function is ensured by the asymptotic condition expressed through the weak constraint
\begin{equation}
\ln[\rho(x)] \rightarrow \ln[\sigma(x)]\,,
\end{equation}
for $x \rightarrow \infty$.

Using the substitution shown in~\ref{q-Boltzmann-factor} we get the generalized version of the partition functions,
\begin{equation}
 Z_q(V_o,T)=\int_0^{\infty}\sigma(E)[1+(q-1)\beta E]^{-\frac{q}{(q-1)}} dE
 \label{Zq1}
 \end{equation}
 and
 \begin{equation}
 \ln[1+Z_q(V_o,T)]=\frac{V_o}{2\pi^2}\sum_{n=1}^{\infty}\frac{1}{n}\int_0^{\infty}dm \int_0^{\infty}dp \, p^2 \rho(n;m)[1+(q-1)\beta \sqrt{p^2+m^2}]^{-\frac{nq}{(q-1)}} \,,
\label{Zq2}
\end{equation}
respectively. Since experiments have shown that $q>1$, this restriction is adopted in what follows.

We will develop a self-consistent theory where these two definitions must be asymptotically equivalent. In the following we use a few times the transformation
\begin{equation}
[1+(q-1)ax]^ {-\frac{1}{q-1}} \rightarrow [1+(q'-1)x]^ {-\frac{a}{q'-1}}\,,
\end{equation}
with $(q-1)a=(q'-1)$. In all cases this transformation is used for calculation purposes only. We start studying the integral
\begin{equation} I_n(m)=\int_0^{\infty}dp \, p^2 [1+(q-1)\beta \sqrt{p^2+m^2}]^{-\frac{nq}{(q-1)}} \,,
\end{equation}
which we write in the form
\begin{equation}
 I_n(m)=m^3 \int_1^{\infty}dx \, x\sqrt{x^2-1} \, [1+(q'-1) x]^{-\frac{n\beta m + n(q'-1)} {(q'-1)}} \,.
\end{equation}
with $q'$ such that $(q-1)\beta m=q'-1$,

Observe that for $n\beta m \rightarrow \infty$ the main contribution to the integral comes from values close to $\bar{x}=\sqrt{1+\frac{1}{n\beta m}}$. Linearizing the function $g(x)=x\sqrt{x^2-1}$ around $\bar{x}$ we obtain
\begin{equation}
 g(x)\approx \sqrt{n\beta m}(x-1)\,.
 \end{equation}

Using this approximation for  $g(x)$ in the expression for $I_n(m)$ it follows that
\begin{equation}
 I_n(m) \approx m^3\sqrt{n\beta m}\int_1^{\infty}(x-1)[1+(q'-1) x]^{-\frac{n\beta m}{(q'-1)} - n} dx \,.
 \label{expr-In1}
\end{equation}
The integration can be easily performed, resulting
\begin{equation}
\int_1^{\infty}[1+(q'-1) x]^{-\frac{n\beta m+n(q'-1)}{(q'-1)}} dx=\frac{q'^{-\frac{n\beta m+(n-1)(q'-1)}{(q'-1)}}}{n\beta m+(n-1)(q'-1)}
\end{equation}
and
\begin{equation}
 \begin{split}
 &\int_1^{\infty} x [1+(q'-1) x]^{-\frac{n\beta m+n(q'-1)}{(q'-1)}} dx= \frac{ q'^{-\frac{n\beta m+(n-1)(q'-1)}{(q'-1)}} }{n\beta m+(n-1)(q'-1)} + \\
&  \frac{1}{n\beta m+(n-1)(q'-1)}\times \frac{q'^{-\frac{n\beta m+(n-2)(q'-1)}{(q'-1)}}}{n\beta m+(n-2)(q'-1)} \,.
 \end{split}
\end{equation}
For  $(q'-1)=(q-1)\beta m \ll \beta m$ and $\beta m \gg 1$ we get
\begin{equation}
 I_n(m)=\frac{m^{3/2}}{(n \beta)^{3/2}} \big[1+(q-1) n \beta m\big]^{-\frac{1}{q-1}} \,,
\end{equation}
where the relation between $q'$ and $q$ was used.

Substituting $I_n(m)$ into Equation~\ref{Zq2} for $Z_q(V_o,T)$ it results that
\begin{equation}
Z_q(V_o,T)=\exp\bigg\{\frac{V_o}{2\pi^2 (n \beta)^{3/2}}\sum_{n=1}^{\infty}\frac{1}{n}\int_0^{\infty}dm\, m^{3/2} \rho(n;m) \big[1+(q-1)n \beta m \big]^{-\frac{1}{q-1}}\bigg\}-1
\end{equation}
Clearly the leading term corresponds to $n=1$, therefore we will drop the summation in what follows.

According to the asymptotic bootstrap principle, for $m,E \rightarrow \infty$ the two expression for $Z_q$ must be equivalent, thus
\begin{equation}
Z_q(V_o,T)= \int_0^{\infty}\sigma(E)[1+(q-1)\beta E]^{-\frac{q}{(q-1)}} dE \\  \nonumber
          =  \exp\bigg\{\frac{V_o}{2\pi^2 \beta^{3/2}} \int_0^{\infty}dm\, m^{3/2} \rho(m)  [1+(q-1)\beta m]^{-\frac{1}{q-1}}\bigg\}-1 \,,
\end{equation}
with $\rho(m)=\rho(1;m)$. At the same time the weak constraint on the mass and energy densities,
\begin{equation}
 \ln[\sigma(E)]=\ln[\rho(m)] \,,
\end{equation}
must be fullfilled.

Now we show that the self-consistency can be asymptotically achieved by choosing
\begin{equation} m^{3/2} \rho(m)=\frac{\gamma}{m}\big[1+(q_o-1) \beta _o m\big]^{\frac{1}{q_o -1}}=\frac{\gamma}{m}[1+(q'_o-1)  m]^{\frac{\beta _o}{q'_o -1}}
\end{equation}
and
\begin{equation}
\sigma(E)=bE^a\big[1+(q'_o-1)E\big]^{\frac{\beta _o}{q'_o -1}}\,,
\label{q-level-density}
 \end{equation}
where $\gamma$ is a constant and $q'_o-1=\beta _o (q_o-1)$. Here, $a$, $b$ and $\gamma$ are arbitrary constants. Note that for $q' \rightarrow 1$ the two expressions above approach the corresponding expressions in the Hagedorn's theory.

Using the above expression for $\rho(m)$ in that for $Z_q$ we get
\begin{equation}
\begin{split}
1+Z_q(V_o,T)= & \exp\bigg\{\frac{V_o}{2\pi^2 \beta^{3/2}}\bigg[\int_0^M dm \, m^{3/2} \rho(m)\big[1+(q''-1)m\big]^{-\frac{\beta}{q''-1}}+ \nonumber \\
 & \int_M^{\infty} \frac{dm}{m} [1+(q'_o-1)m]^{\frac{\beta_o}{q'_o-1}} \gamma [1+(q''-1)m]^{-\frac{\beta}{q''-1}} \bigg]\bigg\}\,,
 \end{split}
\end{equation}
where $M$ has to be chosen sufficiently large and $(q''-1)=\beta(q-1)$. Then
\begin{equation}
1+Z_q(V_o,T)= Z_1(V_o,T)+\exp\bigg\{\frac{\gamma V_o}{2\pi^2 \beta^{3/2}}\int_M^{\infty} \frac{dm}{m} [1+(q'_o-1)m]^{\frac{\beta_o}{q'_o-1}}[1+(q''-1)m]^{-\frac{\beta}{q''-1}}\bigg\}
\end{equation}
with
\begin{equation}
Z_1(V_o,M)=\int_0^M dm \, m^{3/2} \rho(m)\big[1+(q''-1)m\big]^{-\frac{\beta}{q''-1}}\,.
\end{equation}

We can choose $q'_o$ such that $q'_o-1=q''-1$, then
\begin{equation}
 1+Z_q(V_o,T)=Z_1(V_o,M)+\exp\bigg\{\frac{\gamma V_o}{2\pi^2 \beta^{3/2}}\int_M^{\infty} \frac{dm}{m} [1+(q'_o-1)m]^{-\frac{\beta-\beta_o}{q'_o-1}}\bigg\} \,.
 \label{preZq2}
\end{equation}

%
In order to evaluate this integral, note that
\begin{equation}
\int_a^ {\tilde{a}}[1+(q-1)a'm]^{-\frac{1}{q-1}-1} da' =\frac{1}{m}\big\{[1+(q-1)am]^{-\frac{1}{q-1}}-[1+(q-1)\tilde{a}m]^{-\frac{1}{q-1}}\,.
\end{equation}
For any $m>0$ we can find $\tilde{a}>a$ such that
\begin{equation}
 [1+(q-1)am]^{-\frac{1}{q-1}}\gg [1+(q-1)\tilde{a}m]^{-\frac{1}{q-1}}\,,
\end{equation}
so that
\begin{equation}
 \int_a^ {\tilde{a}}[1+(q-1)a'm]^{-\frac{1}{q-1}-1} da' \approx \frac{[1+(q-1)am]^{-\frac{1}{q-1}}}{m}\,.
\end{equation}

 Using this result in Equation~\ref{preZq2} with $a=\beta-\beta_o$ and $(q''_o-1)a'=q'_o-1$, we can write
\begin{equation}
\int_M^{\infty} \frac{dm}{m} [1+(q'_o-1)m]^{-\frac{a}{q'_o-1}} \approx   \int_a^{\tilde{a}}   \int_{\tilde{m}}^{\infty}  [1+(q''_o-1)a'm]^{-\frac{1}{q''_o-1}-1} dm da' -  \int_{\tilde{m}}^{M} \frac{dm}{m} [1+(q'_o-1)m]^{-\frac{a}{q'_o-1}}\,,
\end{equation}
where $\tilde{m}$ is an arbitrary constant.
Integrating on $m$ in the first term of the equation above we get
\begin{equation}
 \int_M^{\infty} \frac{dm}{m} [1+(q'_o-1)m]^{-\frac{a}{q'_o-1}} \approx  \int_a^{\tilde{a}} \frac{[1+(q''_o-1)a'\tilde{m}]^{-\frac{1}{q''_o-1}}}{a'} da' - \int_{\tilde{m}}^{M} \frac{dm}{m} [1+(q'_o-1)m]^{-\frac{\beta-\beta_o}{q'_o-1}}\,.
\end{equation}

Now we can choose $\tilde{m}$ such that
\begin{equation}
 [1+(q''-1)\tilde{a}\tilde{m}]^{-\frac{1}{q-1}} \approx 1\,,
\end{equation}
so that the integration on $a'$ can be easily performed resulting
\begin{equation}
 \int_M^{\infty} \frac{dm}{m} [1+(q'_o-1)m]^{-\frac{\beta-\beta_o}{q'_o-1}} \approx \ln\bigg(\frac{1}{\beta-\beta_o}\bigg)+\ln{\tilde{a}} - \int_{\tilde{m}}^{M} \frac{dm}{m} [1+(q'_o-1)m]^{-\frac{\beta-\beta_o}{q'_o-1}}\,,
\end{equation}
where we substituted $a$ by $\beta-\beta_o$\footnote{An alternative way to obtain this result is noticing that under the conditions relevant here we have
\begin{equation} \nonumber
\int_M^{\infty} \frac{dm}{m} [1+(q-1)am]^{-1/(q-1)} \approx _2F_1\bigg(\frac{1}{q-1},\frac{1}{q-1};1+\frac{1}{q-1};\frac{-1}{aM(q-1)}\bigg)
\bigg(\frac{1}{aM(q-1)}\bigg)^{1/(q-1)}(q-1)\,,
\end{equation}
where $_2F_1$ is the Hypergeometric function.
Also
\begin{equation}\nonumber
_2F_1\bigg(\frac{1}{q-1},\frac{1}{q-1};1+\frac{1}{q-1};\frac{-1}{aM(q-1)}\bigg) \approx \frac{\frac{1}{(q-1)} \ln\bigg(\frac{1}{aM(q-1)}\bigg)}{\bigg(\frac{1}{aM(q-1)}\bigg)^\frac{1}{q-1}}\,,
\end{equation}
resulting
\begin{equation}\nonumber
\int_M^{\infty} \frac{dm}{m} [1+(q-1)am]^{-1/(q-1)} \approx \ln(1/a)-\ln[(q-1)M]\,,
\end{equation}
which is equivalent to the result derived here.
}

With this result and Equation~\ref{preZq2} we obtain the asymptotic form of the partition function
\begin{equation}
 Z_q(V_o,T) \rightarrow \bigg(\frac{1}{\beta - \beta _o }\bigg)^{\alpha}+F_q(V_o,M)-1 \label{asymptZq}
 \end{equation}
where
\begin{equation}
\alpha=\frac{\gamma V_o}{2\pi^2 \beta^{3/2}}\,,
 \end{equation}
and
\begin{equation}
F_q(V_o,M)=Z_1(V_o,M)+ \exp\bigg\{\frac{\gamma V_o}{2\pi^2 \beta^{3/2}}\bigg[\ln{\tilde{a}}-\int_{\tilde{m}}^{M} \frac{dm}{m} [1+(q'_o-1)m]^{-\frac{\beta-\beta_o}{q'_o-1}}\bigg] \bigg\} \label{eqFq}\,.
\end{equation}
Observe that $F_q(V_o,M)$ approaches a constant as $\beta \rightarrow \beta _o$.

Now we turn our attention to the second expression for $Z_q$, presented in Equation~ \ref{Zq1}, that is,
\begin{equation}
Z_q(V_o,T)=\int_0^{\infty} b E^{a}[1+(q'_o-1)E]^{-\frac{\beta -\beta _o}{q'_o-1}-1} dE \,,
\label{Zq1-2}
\end{equation}
where Equation~ \ref{Zq2} was used.

The right side of this equation is the q-Laplace Transform of the function $h(E)=b E^{\alpha}$, according to the definition given by~\cite{Borges}, which results to be
\begin{equation}
Z_q(V_o,T)=b \Gamma(a+1) \frac{\Gamma \bigg(\frac{\beta-\beta _o}{q'_o-1}-a\bigg)}{(q'_o-1)^{a+1} \Gamma \bigg(\frac{\beta-\beta _o}{q'_o-1}+1\bigg)}
\end{equation}
Here we have the constraint
\begin{equation}
\frac{\beta-\beta _o}{q'_o-1}>a+1
\end{equation}
to have convergence in integration of Equation~\ref{Zq1-2} .

Using the properties of the $\Gamma(z)$ function it follows that for $(q'_o-1) \rightarrow 0$
\begin{equation}
Z_q(V_o,T) \rightarrow b \Gamma(a+1)\bigg( \frac{1}{\beta-\beta _o}\bigg)^{a+1}\,.
\label{asymptotic_Zq}
\end{equation}
Therefore we can make the two expressions for $Z_q$ to converge if
\begin{equation}
 a+1=\alpha=\frac{\gamma V_o}{2\pi^2 \beta^{3/2}} \,.
\end{equation}

One immediate consequence of equation~\ref{asymptotic_Zq} is the existence of a limiting temperature $T_o=1/\beta _o$. This result is equivalent to the important result obtained by Hagedorn, and it was already obtained for the case of non-extensive statistics by Biró and Peshier~\cite{Biro}


The existence of such a limiting temperature can be investigated through the fitting to experimental data of the transverse momentum ($p_{\perp}$) distribution\cite{Bediaga2000156,Beck}
\begin{equation}
\frac{1}{\sigma}\frac{d\sigma}{d p_{\perp}}=c[2(q-1)]^{-1/2}B\bigg(\frac{1}{2},\frac{q}{q-1}-\frac{1}{2}\bigg)u^{3/2}[1+(q-1)u]^{-\frac{q}{q-1}+\frac{1}{2}}\,,
\label{pt}
\end{equation}
where $u=\beta_o p_{\perp}$, and $B(x,y)$ is the Beta-function. According to the analysis performed in Reference~\cite{Bediaga2000156}, the effective temperature obtained for center of mass energy in the range from 35 Gev up to 160 GeV is approximately constant around $T_o \sim$ 110 MeV. Therefore, the constant temperature obtained here is in accordance with experiments. Also the entropic parameter $q$ is approximately constant around $q \sim $ 1.2 in the energy range analyzed. 

In~\cite{Aguiar2003371} the authors argue that there would not be a relativistic ideal gas in non-extensive thermodynamics because the partition function presents a singularity for $\beta \rightarrow \infty$. The result obtained here shows that this limit is never achieved, and therefore the multiplicity distribution derived in \cite{Aguiar2003371} could be used to determine the hadron multiplicity in high energy strong interactions.

Finally, it is important to note that the average occupation number which is used in~\cite{Bediaga2000156,Beck} and implicitly used in the present work, is consistent with the thermodynamical relations in the sense discussed in Ref.~\cite{Conroy} at the limits $m \rightarrow \infty$ and $E \rightarrow \infty$.

After the present work was completed the author was informed about a recente analysis of the effective temperature and of the entropic parameter in $pp$ collision at 900 GeV~\cite{Cleyman_Worku}. In this work the authors analyzed the transversal momentum distributions for many different particles produced in the collision, and showed that all distributions can be well described with $T\approx$70 MeV and $q\approx$ 1.15, confirming the findings obtained with the self-consistency study performed here.  The different values obtained by Bediaga~\cite{Bediaga2000156} and by Cleyman and Worku~\cite{Cleyman_Worku} can be attributed to the slightly different formulae used in each of these works. In the last one the authors used an occupation number formula which is consistent with the thermodynamical relations, as discussed above. It is important to emphasize that here $T$ refers to the effective temperature and can only be compared to the Hagedorn`s temperature through a linear relation, as described in Refs~\cite{Wilk2007,Wilk2009}

The existence of a constant temperature, $T_o$, and a constant entropic factor, $q_o$, imposes strong constraints on the applicability of the non-extensive statistics to ultra-relativistic collisions. In this way, the present work gives an interesting tool to investigate the role of non-extensive statistics in strongly interacting systems. 

Concluding, in this work  we have shown that
\begin{enumerate}
\item It is possible to generalize the self-consistency principle using the non-extensive statistics.
\item When substituting $\exp\{-\beta E \}$ by $[1+(q-1)E]^{-q/(q-1)}$ in the partition function we have to adopt
\begin{equation}
\rho(m)  \rightarrow \gamma m^{-5/2}[1+(q'_o-1)  m]^{\frac{\beta _o}{q'_o -1}}
\end{equation}
and
\begin{equation}
  \sigma(E) \rightarrow bE^a\big[1+(q'_o-1)E\big]^{\frac{\beta _o}{q'_o -1}}
\end{equation}
 in order to get a self-consistent theory.

\item With these choices we get the following asymptotic form for the partition function:
\begin{equation}
 Z_q(V_o,T) \rightarrow \bigg( \frac{1}{\beta-\beta _o}\bigg)^{a+1}\,.
 \end{equation}
\item From here it follows that there is an inferior limit for $\beta$ corresponding to a limiting effective temperature $T_o=1/\beta _o$.
\end{enumerate}

The author thanks Dr. E. P. Borges for carefully reading the manuscript and for the valuable suggestions. This work was supported by the Brazilian agency, CNPq, under grant 305639/2010-2 .


\end{document}